# A NORMATIVE APPROACH TO RADICALIZATION IN SOCIAL NETWORKS

Vincent BOUTTIER[1,2], Salomé LECLERCQ[2], Renaud JARDRI[1,2,*,#], Sophie DENÈVE[1,#]

1. PSL University, Laboratoire de Neurosciences Cognitives et Computationnelles (LNC²), INSERM U960, DEC, France
2. Lille University, Centre Lille Neurosciences & Cognition, INSERM U1172, CHU Lille, Plasticity & SubjectivitY team, France

* correspondence to : renaud.jardri@univ-lille.fr ; phone: +33 (0) 320 446 747
# equal contributions

**ABSTRACT** (155 words):

In recent decades, the massification of online social connections has made information globally accessible in a matter of seconds. Unfortunately, this has been accompanied by a dramatic surge in extreme opinions, without a clear solution in sight. Using a model performing probabilistic inference in large-scale loopy graphs through exchange of messages between nodes, we show how circularity in the social graph directly leads to radicalization and the polarization of opinions. We demonstrate that these detrimental effects could be avoided if the correlations between incoming messages could be decreased. This approach is based on an extension of Belief Propagation (BP) named *Circular Belief Propagation* (CBP) that can be trained to drastically improve inference within a cyclic graph. CBP was benchmarked using data from Facebook© and Twitter©. This approach could inspire new methods for preventing the viral spreading and amplification of misinformation online, improving the capacity of social networks to share knowledge globally without resorting to censorship.

**ACKNOWLEDGMENTS AND FUNDING SOURCES**:

**Author contributions**: VB generated the graphs and performed the analyses under the supervision of RJ and SD. All the authors contributed equally to discussion and writing the manuscript.

**Funding**: VB was supported an ANR grant attributed to RJ (INTRUDE ANR-16-CE37-0015) and by the Fondation pour la Recherche Médicale (FRM grant FDT02001011086).

**Competing interests**: Our funding sources had no role in the design of the study and did not have any role during its execution, analyses, interpretation of the data, or decision to submit results.

**Data availability Statement**: No data associated in the manuscript.

# **INTRODUCTION**

Online social networks have great benefits and advantages. They allow for the quasi-instantaneous exchange of up-to-date information and give access to persons around the world with different backgrounds, experiences and opinions. They also create communities with sizes well beyond the usual social constraints, and perhaps even beyond cognitive ones [1]. Nevertheless, the constant increase in network size and complexity may introduce more information than we can normally process [2], as well as promoting passionate (and sometimes extreme) debates. Beyond the initial excitement these networks provided, the regular polarization of positions on social media appears worrisome. For example, it promotes severe conflicts between communities expressing opposite beliefs, while also making social networks particularly vulnerable to manipulation or propaganda, for instance, by bots accused of interference with presidential elections [3].

Solutions need to be found, but without sacrificing the advantages of worldwide information access or impoverishing social interactions. In our view, the problem goes far beyond the propagation of fake news, which is a symptom as much as a cause of polarization. More than the content of one's belief, the issue seems to revolve around overconfidence and excessive trust (or distrust) in information confirming (or contradicting) these beliefs. Many of the most polarizing issues discussed on the internet may not even have a universally defined, knowable, or absolute answer (this is the case for societal questions such as immigration policies but also questions beyond these such as the existence of extraterrestrial intelligence). For these issues, *radicalization* can be defined as people reaching unreasonably confident and monolithic beliefs based on multifaceted, biased or untrustworthy data [4]. Additionally, the emergence of two or more radicalized groups with opposite, irreconcilable beliefs results in *polarization*.

To capture these phenomena in a simplified, mathematically grounded but intuitive framework, we treat large-scale opinion sharing in social networks as a form of probabilistic inference. People's beliefs are modeled as the probability of giving an answer to a particular question (e.g. *Should abortion be legal or not?*). Rather than just deciding "yes" or "no" once for all (a binary choice), someone could have a graded confidence level represented with a probability, close to 100% or 0% for high confidence or equivalently strong opinions, but approaching 50% if the person is uncertain. Agents embedded in a social network derive their beliefs both from external or private sources of evidence (direct experience, expertise, news articles, religious values, etc.) and from the expressed opinions of people they are connected to or communicate with (see **Fig.1a**). Through communication, that is, the propagation of messages within a social network, each person's opinion should ideally become as informed as possible, integrating the knowledge and experience from all the network members. In other words, we work under the hypothesis of normativity, according to which the purpose of communication is to ensure that individual opinions converge to a consensus corresponding to the posterior probability of the answer given all the external evidence. This "ideal" situation, well defined mathematically, represents a benchmark against which various message propagation schemes can be compared, while significant deviation (such as systematic overconfidence) could be considered irrational.

Unfortunately, the structure of social networks renders simple message passing schemes fatally flawed as an inference mechanism (see **Fig.1b-d**). In particular, every loop in a social graph forms an *echo chamber* where opinions can reverberate *ad infinitum* and be artificially amplified [5,6] (see **Fig.1d,e**). We thus confront both the strengths and weaknesses of the massification of social media: social networks could (ideally) make local information globally available as never before. However, they also tend to aberrantly amplify confidence, leading to radicalization and polarization and, as we will see, severely limiting their true information sharing capability.

The goal of this paper is twofold. First, we provide a simple account of echo chambers using a probabilistic inference framework (BP) applied to realistic social graphs and systematically study their consequences. Second, we propose a method (CBP) that limits these detrimental effects by trying to achieve normality, bringing the confidence levels generated in the network closer to informed rationality. We demonstrate the efficiency of this algorithm in both toy graph-models and more realistic graph structures borrowed from popular social networks.

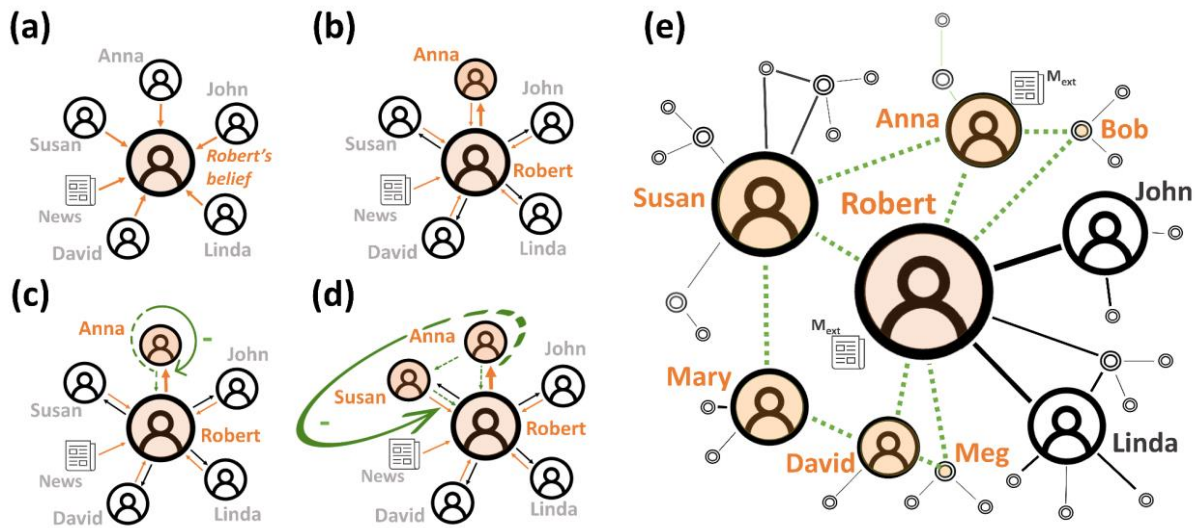

**Figure 1. A normative account of message-passing in social networks. (a)** Let us consider a social network in which an agent, for instance Robert, communicates with other agents one-to-one and reads the news. In this context, Robert's belief can be computed as the sum of all the information he receives: internal messages from other agents and external messages ($M_{ext}$). **(b)** What information should Robert communicate to Anna? An initial answer could be all the information Robert received, including information that came from Anna (the *mean-field implementation*). This approach is suboptimal since redundant information is exchanged and thus counted several times. **(c)** To address this overcounting, a second implementation (*Belief Propagation*) would consist of sending Anna all the information received by Robert except the information coming from her (dashed green arrow). This message cancellation is indicated by the green arrow. **(d)** However, the problem becomes much more complex when Anna and Robert have common friends like Susan. In this case, Robert's belief is corrupted by what Susan knows from Anna. Therefore, an extra correction has to be applied to control the flow of circular messages in the social graph. The *Circular Belief Propagation* (CBP) algorithm implements such a correction. **(e)** Without proper control, the highlighted problem becomes much more serious when the social graph is highly cyclic, when adding new friends/followers such as Mary, Meg and Bob (see green connections).

# **RESULTS**

1 - A BRIEF OVERVIEW OF MESSAGE PASSING SCHEMES

In our simplified social network model, each agent is a node of the network, and edges of the network represent the social circle of agents (the people they directly communicate with, that is, friends or followers). We assume that each agent *i* estimates the true probability distribution of a binary variable $x_i$ ($x_i$ = "yes" or "no", for instance to the question *Should abortion be legal or not?*), given diverse sources of external information (web-search, articles, books, TV programs, or even direct evidence such as that coming from field journalists or researchers). In the following, we will refer to $p(x_i = \text{"yes"})$ as the *true probability* (formally the true *marginal posterior probability* of $x_i$ to be "yes") and to $b(x_i = \text{"yes"})$ as the *probability estimate*. For convenience, we also define the *belief* $B_i$ of agent *i* (that $x_i$ is "yes" versus "no") such that $b(x_i = \text{"yes"}) = \text{sigmoid}(2B_i) = 1/(1+e^{-2B_i})$; see **Sup.Fig.1**.

The sign of $B_i$ describes the agent's opinion about the binary question: if for example $B_i > 0$, the agent believes that the answer to the corresponding question is more likely to be "yes" than "no". Additionally, the absolute value |B$_i$| quantifies the *confidence* of agent $i$. The higher the confidence, the more certain the agent is about the answer, while $|B_i| = 0$ implies complete uncertainty.

To determine the value of his/her belief $B_i$, the agent has to combine two types of information:

- External sources of information received by this agent, or agent's preference, grouped together and quantified as the *external message* $M_{\text{ext} \to i}$. Such a message (mathematically defined as a log-likelihood ratio) is negative if it supports "no", and positive if it supports "yes" ; The amplitude of this external message indicates its assumed reliability.

- Information provided by the opinions broadcasted by members of the agent's social circle (called *internal messages* in the following). $M_{i \to j}$ denotes the message sent from agent i to agent j.

An agent's core belief is the sum of all the internal and external messages it receives (see **Fig.1a**):

$$B_i^t = \sum_j M_{j \to i}^t + M_{\text{ext} \to i}$$

(Equation 1)

Meanwhile, the message $M_{j \to i}$ depends on the belief of agent j, and the amount of trust [7] between the two agents i and j. In the simplest possible message passing scheme, called *variational message-passing* [8], the message corresponds to the sender's belief modulated by trust: $M_{j \to i}^{t+1} = f_{ji}(B_j^t)$, where $f_{ij} = f_{ji}$ is a sigmoidal function which depends on the (mutual) amount of trust between the two agents. This naive method of communication assumes that agents systematically broadcast their opinion to their entire social circle, and in turn combine internal and external messages to update their own beliefs (see **Fig.1b**). This message-passing algorithm corresponds to what was proposed in previous models of opinion dynamics in social networks [5,9,10] with slight differences in the precise form of the sigmoidal function (see **Methods**). However, the above mean-field scheme is highly suboptimal at performing inference in a graph. In particular, it creates a (potentially uncontrolled) reverberation of messages between each connected pair of nodes: agent j influences i, who influences j, etc. Humans probably never communicate this way; for instance, we only tell our friends things they presumably do not already know.

A less naive communication method, which we hypothesize to be our model for communication, ensures that messages do not include the messages sent previously in the opposite direction (see **Fig.1c**). Messages are updated iteratively as follows: $M_{j \to i}^{t+1} = f_{ji}(B_j^t - M_{i \to j}^t)$. The resulting message passing scheme corresponds to a widely used inference algorithm called *Belief Propagation* (BP) [11] (see **Math Appendix**). Despite its simplicity, this algorithm is surprisingly powerful as an (approximate) inference method [12]. In fact, BP is even exact in graphs without cycles, that is, it converges to the true posterior probabilities. However, in the presence of cycles, messages can still be reverberated and artificially amplified, leading to overconfidence, shown schematically in **Fig.1d**. Unfortunately, social networks contain a large number of such loops (see **Fig.1e**). As a result, we will see that BP, considered as a model of social communication, systematically leads to radicalization and polarization in cyclic social graphs.

As a society, we urgently need to find solutions that can preserve the global knowledge sharing capabilities of social networks, while suppressing the detrimental effects of loops or echo chambers. When integrating information from someone, one should in theory consider all the indirect ways the content has been brought to him or her (through a common friend for instance) in order to not take into account the same piece of information twice. With this goal in mind, we introduce an adaptation of the Belief Propagation algorithm called *Circular Belief Propagation* (CBP) [13] which aims at actively removing redundancies between messages introduced by loops and amplification of messages through cycles. The resulting message passing scheme can be written as follows:

$$M_{j \to i}^{t+1} = f_{ji}(B_j^t - \alpha_{ij} M_{i \to j}^t)$$ (Equation 2)

where beliefs are defined by:

$$B_i^t = \kappa_i \left( \sum_j M_{j \to i}^t + M_{\text{ext} \to i} \right)$$ (Equation 3)

In contrast to BP, CBP contains two types of control parameters: a gain $\kappa_i$ applied to each node, and a loop correction term $\alpha_{ij} = \alpha_{ji}$ applied to each link. The idea of the first equation is to subtract more than once the opposite message $M_{i \to j}$ from the belief of agent j. This is based on the fact that agent j is not only influenced directly by i, but also indirectly by any person k (all messages $M_{k \to j}$ might contain some part of $M_{i \to j}$ as i might influence k). Intuitively speaking, the loop correction term "$-\alpha_{ij} M_{i \to j}$" subtracts the predictable "redundant" part from incoming messages, which is the result of the reverberation of the outgoing message through all the graph's loops. Similarly, the gain $\kappa_i$ in the second equation prevents the amplification of beliefs due to excess correlations between all incoming messages as introduced by loops (agents influence themselves, as messages travel back).

Note that these control parameters would need to be adapted to the local structure around each node within the graph, thus posing an additional challenge. Here we will consider two methods of finding a good set of control parameters : (a) a supervised learning method, that can only be used in extremely small graphs, and (b) a local unsupervised learning rule that is less optimal but applicable to graphs of arbitrary sizes (see **_Methods_**). All control parameters can be trained in an unsupervised manner by ensuring that incoming and outgoing messages remain as decorrelated as possible when they contain no meaningful information.

To model opinion formation in a social network, we iterate 100 times the BP/CBP message passing scheme simultaneously in all the nodes, to let the information provided by the external messages propagate in the entire graph (at which stage beliefs and messages usually reach a stable state). Further details are provided in the **_Methods_** section and the pseudo-code is given in the **_Math Appendix_**.

2 - PERFORMANCE OF THE BP AND CBP ALGORITHMS

**Figure 2**. **Performance of message passing algorithms in 10-node toy examples**.

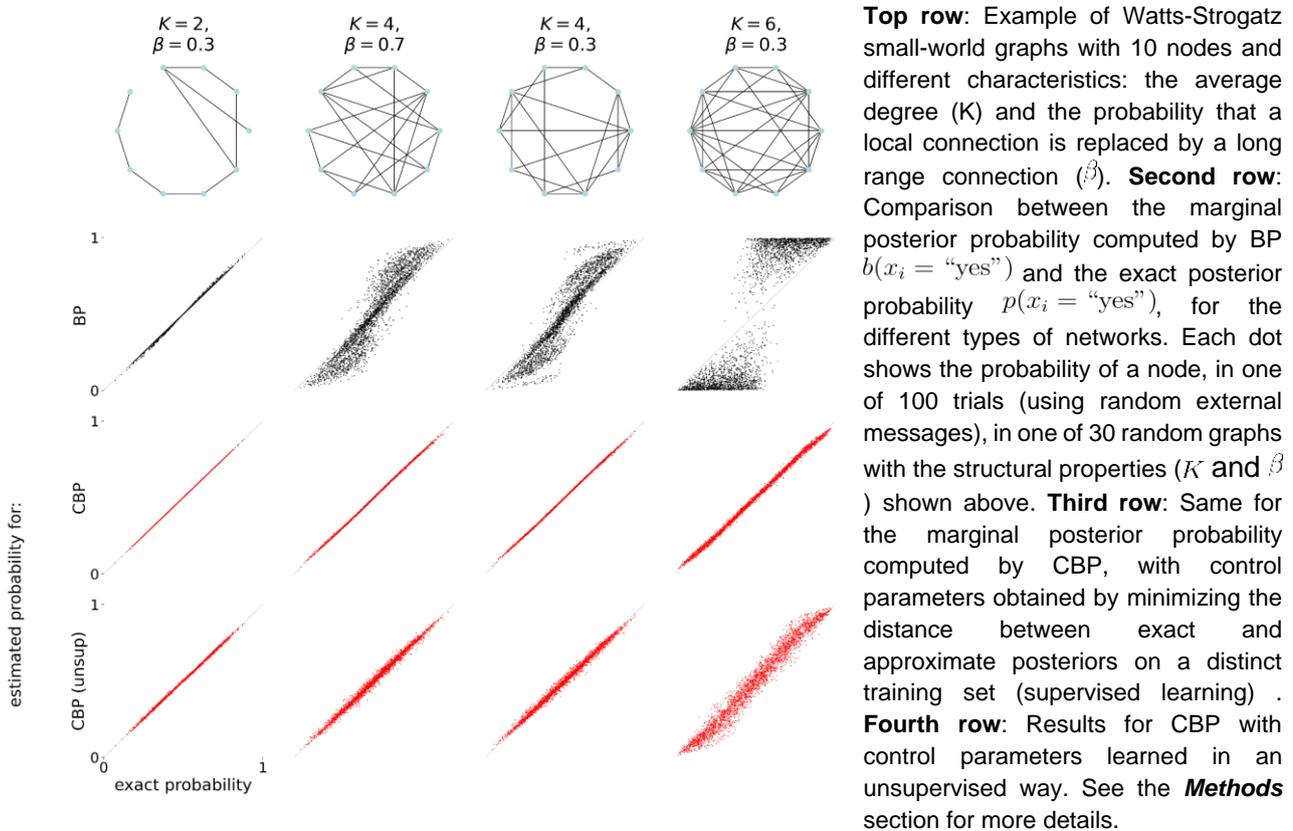

**Top row**: Example of Watts-Strogatz small-world graphs with 10 nodes and different characteristics: the average degree (K) and the probability that a local connection is replaced by a long range connection ($\beta$). **Second row**: Comparison between the marginal posterior probability computed by BP $b(x_i = \text{"yes"})$ and the exact posterior probability $p(x_i = \text{"yes"})$, for the different types of networks. Each dot shows the probability of a node, in one of 100 trials (using random external messages), in one of 30 random graphs with the structural properties ($K$ and $\beta$) shown above. **Third row**: Same for the marginal posterior probability computed by CBP, with control parameters obtained by minimizing the distance between exact and approximate posteriors on a distinct training set (supervised learning) . **Fourth row**: Results for CBP with control parameters learned in an unsupervised way. See the **_Methods_** section for more details.

We first tested the performance of BP and CBP in small graphs (toy examples with $n = 10$ nodes) where running exact probabilistic inference is still practical, as well as supervised learning (see **Fig.2**). This way, the resulting exact posterior marginal probabilities can be used as a benchmark (an ideal result) for comparison with BP or CBP. The graphs were generated to have a Watts-Strogatz *small-world* structure, as the latter share some features with social networks [14]. Such a graph structure is controlled by two parameters: the mean degree (number of connections) for each node, called $\kappa$, and the probability that a connection is "long range" as opposed to local between neighbors, called $\beta$ (see **Fig.2a** for example structures).

We compared the approximate posterior probability solutions $b(x_i = \text{"yes"})$ found by different message passing schemes with the exact posteriors, $p(x_i = \text{"yes"})$, for given external messages $M_{\text{ext}\to 1}, M_{\text{ext}\to 2}, \ldots, M_{\text{ext}\to n}$. To ensure the generality of the results, this comparison was performed for several randomly generated graph structures (30 graphs for each setting of the structural parameters) and in response to numerous sets of randomly generated external messages ("trials"). Each dot in **Fig.2** corresponds to an approximate posterior probability for one node in a given trial. As the density of the network increased, the performance of BP degraded (**Fig.2b**). In particular, all beliefs became too extreme, resulting in a condensation of approximate posteriors close to 100% or 0% even when the external evidence did not justify such confidence (for instance if the true posterior was in fact close to 50%).

Next, we tested CBP after learning the control parameters using a supervised learning method (provided in **_Methods_**). The parameters $\kappa_i$ and $\alpha_{ij}$ were chosen to minimize the distance between the approximate and true posterior (on a training set independent from the test set shown in the figure). After optimization, CBP matched the exact inference very closely, with no sign of overconfidence (**Fig.2c**).

Such supervised optimization is only possible in networks with a relatively small number of nodes. In larger networks, and any realistic social graph, exact probabilistic inference is impossible because it scales exponentially with the number of nodes. Fortunately, CBP parameters can also be trained without any knowledge of the true posteriors. Using purely local learning rules, the control parameters can be trained to remove correlations between incoming and outgoing messages and to suppress redundancies between incoming messages (see **_Methods_**). Despite the heuristic nature of these learning rules, the approximate posteriors remain close matches to the true posteriors (**Fig.2e**).

This toy example demonstrates that CBP can alleviate the overconfidence problem associated with BP in cyclic graphs, resulting in more rational beliefs. Since we move on in the next section to larger graphs where exact inference is intractable, it is assumed that the parameters of CBP were trained for each graph structure using the proposed local, unsupervised learning rules, rather than with supervised learning.

3 - TOWARDS GREATER REALISM: LARGER GRAPHS

The next step was to investigate how these effects generalize to more realistic social graph structures. First, we investigated larger (but still simplistic) Watts-Strogatz graphs with 200 nodes. By systematically varying $\kappa$ and $\beta$, we explored the impact of the number of connections per node and long-range connections. These findings will be useful for explaining more complex behavior in "realistic" social graphs (see the next section).

**Fig.3** shows an example graph with moderate degree ($K = 30$) and proportion of long-range connections ($\beta = 0.12$). We provided unreliable external messages that did not strongly support a "yes" or "no" answer. More specifically, in each "trial", each external message was sampled from a Gaussian distribution with zero mean and standard deviation $\sigma_{\text{ext}} = 0.1$.

To understand how opinions are formed, it can be useful to visualize the belief trajectory during the deliberation process, that is, while messages are still being propagated. **Fig.3a**, top row, examines the case of BP. Starting from complete uncertainty ($B_i = 0$ for all agents *i*), the beliefs in the different nodes evolve over the iterations of the BP message passing scheme until they stabilize at constant levels, representing the *opinions* generated by BP. Differences in opinions among the nodes are induced by random variations in local graph structures and in the external messages the nodes receive. Each new trial generates a different set of opinions (left and right panels of **Fig.3a**). Note that the beliefs converge to very large values (either positive or negative), most agents being at least 99% confident in their answer (see **Fig.3c** for the relationship between beliefs and probabilities).

While it is not tractable to compute the exact posteriors, we can estimate an upper bound on "rationality" (the dashed line). This corresponds to the belief of a universal observer summing all the external messages directly: $B_{\text{univ}} = \sum_i M_{\text{ext} \to i}$ [1]. Beliefs larger than $B_{\text{univ}}$ (in absolute value) are necessarily overconfident, since they go beyond the total external evidence. As we can see, BP results in severe overconfidence for most nodes in the graph, despite the true unreliability of external messages. Note that $B_{\text{univ}}$ is an upper bound, not an exact posterior. In fact, if inference were exact, the agents would have significantly lower confidence than the universal observer, for two reasons: the nodes do not trust each other completely (there is always a chance that your friends are wrong…), and not all of them are connected to all other nodes.

In contrast to BP, the CBP algorithm leads to far more moderate opinions (see **Fig.3a,b**, bottom panels), with no sign of radicalization or polarization. The final beliefs are narrowly distributed around a consensus value, which is itself close to zero (low confidence). The beliefs always remain below the universal observer, as would be expected from a rational deliberation process and are in agreement with the completely uninformative nature of the external messages chosen for these trials.

The BP-generated opinions are represented graphically on the top row of **Fig.3b**, illustrating how opinions can be distributed as a function of the proximity (inverse path length) between two nodes. Only two possible outcomes were observed in those graphs. In the first scenario, the entire population reaches the same extreme opinion, either for or against (**top-right panels, Fig.3a,b**). We interpret this phenomenon as a *radicalization* of the entire population. In the second scenario, two populations with opposite but similarly extreme opinions emerge. These populations are separated into 2 or more local clusters within the graph (***top-left panels***, **Fig.3a,b**). We interpret this as *polarization*. We quantify the level of *radicalization R* as the mean absolute value of the beliefs and the level of *polarization P* as their mean standard deviation (computed within a single trial). These definitions correspond to (or are highly similar to) the ones used in other studies [5,9,15,16]. The left panels in **Fig.3a,b** have both high radicalization and high polarization, while right panels have high radicalization but low polarization. Note that the only thing differing between the two panels are the external messages (two sets sampled from the same distribution).

---

[1] In practice, this would correspond to the beliefs of all nodes if exact inference were performed in a network with full connectivity and infinite trust (in which case, all the $x_i$ values would collapse to a single binary random variable and all external messages would be noisy evidence for this shared variable). Since our network has limited connectivity and trust, each node can only achieve a lower confidence level, at least if it remains rational.

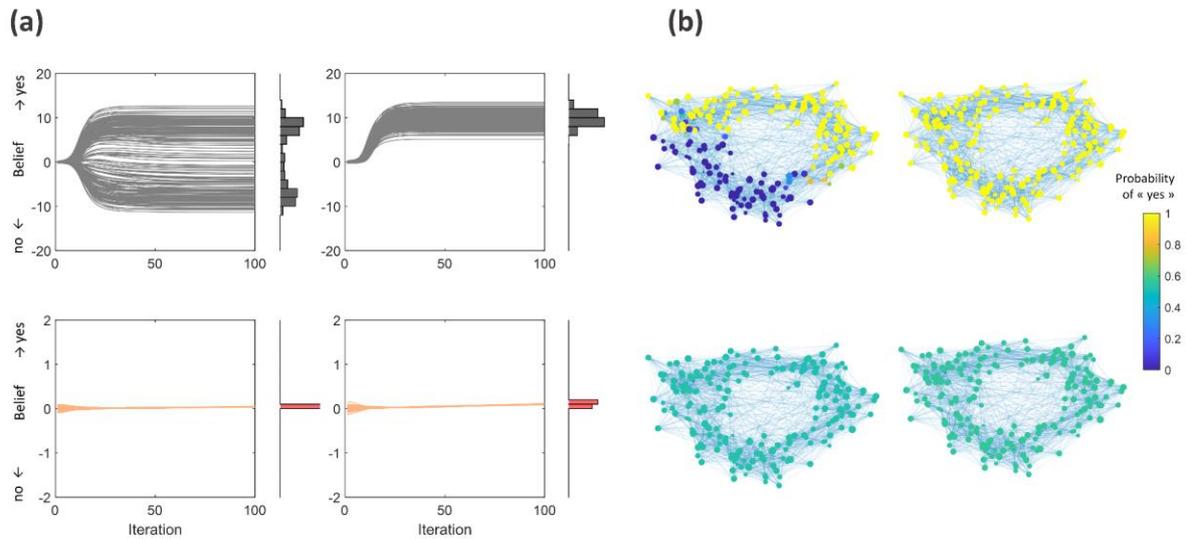

**Figure 3. Response of an example 200-node small-world network ($K$ = 20, $\beta$ = 0.12) to uninformative external messages (random and unbiased)**. **(a)** Temporal evolution of the belief under the message passing algorithm (that is, while internal messages propagate in the network). By convention, the final beliefs are those obtained after 100 iterations. Two example trials are provided. **Top row**: BP leads to two possible outcomes depending on whether the whole population behaves similarly or separates into two groups with opposite, extreme beliefs. This is interpreted as polarization (**left**) or radicalization (unimodal in the **right** panel, bimodal in the **left** panel). **Bottom row**: In contrast, CBP leads to beliefs narrowly distributed around a weak consensus. This consensus varies from trial to trial but remains close to zero, reflecting higher uncertainty. **(b)** Final beliefs of the 200 nodes, visualized in the whole graph. The nodes (dots) are arranged topographically according to the path lengths (separation within the graph). The size of a dot represents the node's degree, and its color represents the marginal posterior probability estimate ($b(x_i = \text{"yes"})$), abbreviated as the "Probability of yes". Thin lines are connections. The two trials shown here are the same as in (a), with the top and bottom rows corresponding to BP/CBP. The relationship between the belief and the "probability of yes" is illustrated in Sup.Fig.1.

The radicalization or polarization due to BP and the suppression of these characteristics by CBP are very general results that are independent of the specific network structure, as illustrated in **Fig.4.** In the case of BP (**Fig.4a**), the severity of polarization and radicalization systematically depends on the two structural parameters: radicalization increases quasi-linearly with $K$ (**left panel**), while polarization decreases with $\beta$ or, equivalently, increases for more clustered networks (right panel). Interestingly, polarization is strongest in a sweet spot with a moderate $K$ and a small value of $\beta$. This sweet spot corresponds to a high probability of echo chambers, which corresponds to local clusters of highly interconnected nodes that are relatively isolated from the rest of the graph (due to the predominance of short-range connections). **Fig.4b** examines in more detail the belief distributions resulting from BP at the level of the population (combined over many trials and several random graphs) when increasing $K$. Note that the distribution has two distinct modes, whose separation increases with $K$.

These features are completely suppressed by CBP. Radicalization and polarization are eliminated (**Fig.4c**), and beliefs are no longer separated into two distinct modes. Instead, the distribution presents a single mode, centered at zero, with a variance increasing with $K$ (**Fig.4d**).

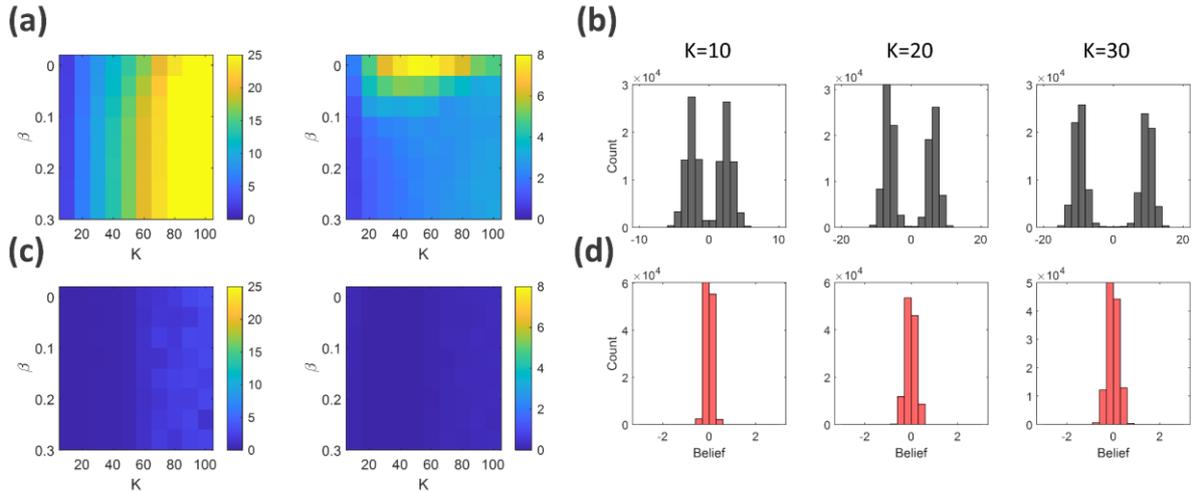

**Figure 4**. **Response of an example 200-node small-world network to uninformative external messages as a function of their structural properties K and beta**. **(a)** Mean radicalization R (left panel) and polarization P (right panel) as a function of $K$ and $\beta$ when using the BP algorithm. **(b)** For BP, the distribution of beliefs over multiple trials, for networks with $\beta$ = 0.12 and increasing values of $K$. **(c)** Same as (a) for CBP. **(d)** Same as (b) for CBP.

Preventing radicalization and polarization is not sufficient per se (for instance, a trivial way of achieving this result would be to set all gains at $\kappa_i = 0$: this way, all beliefs would have been equal to zero). One must also ensure that the message passing scheme operates properly when external messages are actually informative, that is, when they globally provide more support for one option than the other. That is why we now consider a situation in which there is a true answer supported by evidence (such as "*vaccination decreases the risk of severe outcomes following a COVID-19 infection*"). The task of the network is now twofold. First, there should be as many agents as possible whose beliefs point in the direction supported by the evidence (in this case, that most agents believe that COVID vaccines work). Second, confidence levels should increase in proportion to the true strength of this evidence (for instance, skepticism about a therapeutic approach is desirable as long as it has not been validated by rigorous clinical studies).

We generated informative external messages (inputs to the graph) by sampling them from a biased distribution (with a positive mean if the true answer is "yes"). This bias was small compared to the variance of the distribution, such that many individual nodes received misleading external messages (fake news). Moreover, these external messages were injected into only a small portion of the nodes, while others received no external messages (in a situation where the majority of people have no expertise on vaccines). If the network allows all users to share information optimally, every agent should believe in the answers supported by the highest amount of evidence (the sign of the sum of all external messages $B_{\text{univ}}$) even if their private external message points in the opposite direction (that is, even people exposed to fake news would eventually be convinced, through their social contacts, that vaccines are effective).

In investigating inference in the presence of informative messages, we found an interesting dissociation in performance when considering people's choices or their confidence levels. People's *choice* would correspond to their answer to a survey with only two possible options (such as "*Do you think that the COVID vaccine works? yes/no*"). Presumably, they would choose the answer they believe the most, that is, answer "yes" if their belief is positive. In contrast, people's confidence would correspond to the absolute value of their beliefs (for example, "*How confident are you that covid vaccines work/do not work, on a scale from 1 to 10?*").

Let us first consider choices. In a strongly connected network ($K = 40$, $\beta = 0.2$) with small mean path length between nodes (1.9 here), the portion of nodes whose choice is congruent with the external evidence (later called the *congruent* choice) increases similarly to the proportion of informed nodes (receiving external messages). This is true after running either BP or CBP. Moreover, this increase is predicted by a universal observer summing all the external messages together, whose belief is $p(B_{\text{ideal}} > 0)$ (**Fig.5a**, left panel). In a network containing less long range connections ($K = 20$, β = 0.04) with a longer mean path length (2.6 here), both BP and CBP perform worse than the universal observer, reflecting the limitations introduced by the more indirect communication between nodes. However, CBP now clearly outperforms BP (**Fig.5b**, left panel). To intuitively understand why a smaller number of long range connection results in poorer choices, consider an extreme scenario: a network with no long range connections at all ($\beta = 0$), in which case all nodes are organized on a fat ring, with subpopulations at opposite ends having no direct connections. They can only influence each other indirectly by changing the beliefs of intermediate nodes, which is not possible if those nodes are radicalized (as is the case of BP). By keeping beliefs graded, CBP restores long range communication within the network.

Where BP and CBP most strikingly differ is in their confidence levels (that is, answering with not only a yes/no reply but on a scale - **Fig.5a,b**, middle versus right panels). In the case of BP, beliefs are always distributed in two extreme modes, leaving no room for uncertainty (**Fig.5a,b**, right panels). As more evidence arrives in support of a positive choice (more nodes are informed), the proportion of belief in the positive mode (making the "right choice": positive beliefs) increases, but the nodes that are still in the negative mode (making the "wrong choice": negative beliefs) remain equally overconfident (**Fig.5a,b**, right panels). When a node finally changes its mind, it can only switch between these two extremes, with no intermediate stage of uncertainty. Such phenomena could have potentially deleterious societal consequences: people convinced of their correctness could reject the vaccine at any cost and become impervious to information campaigns and contrary evidence; even if they change their minds, one form of extremism could lead to the opposite one. In contrast, with CBP, the beliefs are far less extreme, and their unimodal distribution gradually shifts toward the positive side as more evidence is provided (**Fig.5a,b**, middle panels). In other words, the stronger the evidence, the more confident the congruent agents (nodes with a belief corresponding to the congruent choice) are. Conversely, the incongruent agents become less confident, as should occur following a rational consensus building process.

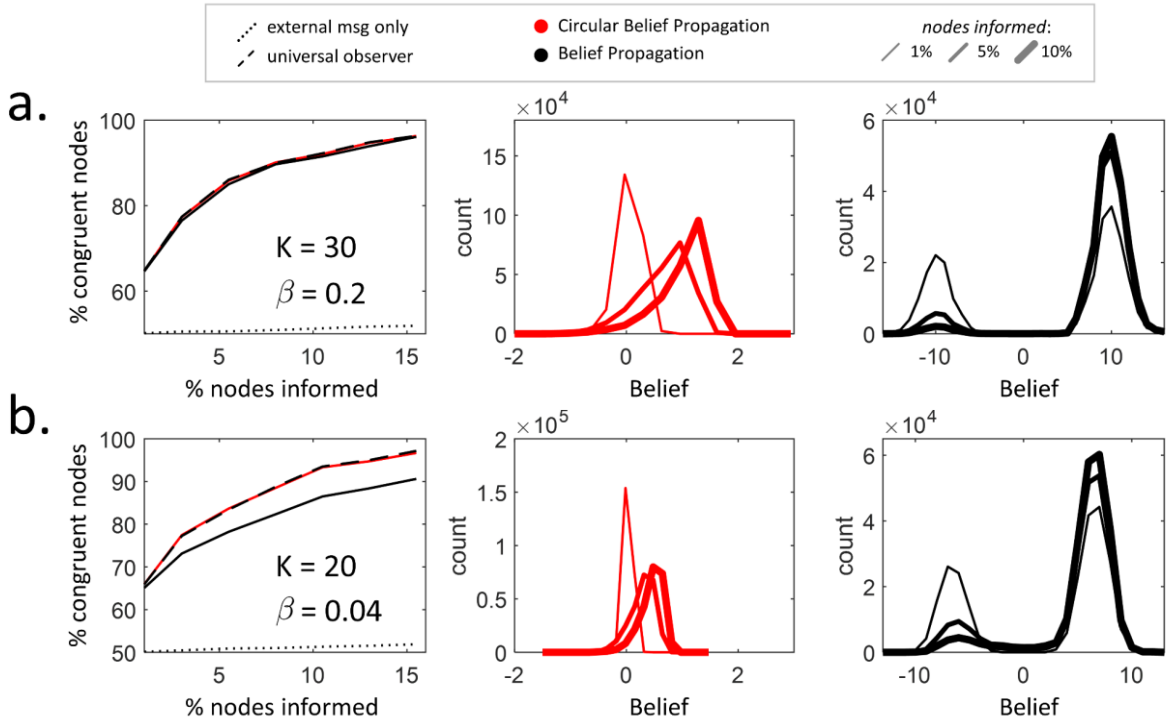

**Figure 5**. **Responses of the 200-node small world graphs to informative (biased) external messages**. **(a)** Average of 6 graphs with $K = 30$, $\beta = 0.2$. **(b)** Same as (a) but for $K = 20$, $\beta = 0.04$. **Left panels**: choice performance of the different message passing schemes. Black: BP, red: CBP, dashed: universal observer, dotted: based on external messages only, without taking into account the internal messages. "% Node informed" is the percentage of nodes receiving non-zero external messages, "% congruent nodes" is the percentage of nodes with a belief whose sign points to the congruent answer (defined by the bias in the distribution of external messages). **Middle panels**: Belief distribution over all nodes for increasing amounts of external information as a result of CBP. Thin line: 1% nodes informed, normal line: 5% of nodes informed, thick line: 10% of nodes informed. **Right panels**: Same as the middle panels, but for BP.

## 4 - REAL SOCIAL-NETWORK EXAMPLES

Finally, we tested BP and CBP on large online social network structures taken from open-access Facebook© and Twitter© data [17] (see **Figs. 6 to 8**). The results are globally consistent with what was observed in toy examples. As before, BP generates aberrantly strong beliefs, even in response to completely uninformative messages. More realistic social graphs contain cliques of highly connected nodes separated by relatively sparse long-range connections. As a result, polarization within local clusters (as opposed to general radicalization of the whole population) was systematically the outcome in response to uninformative external messages (see examples in **Fig 6a,b** top-row). In contrast to BP, CBP generated moderate confidence levels, with no obvious radicalization or polarization (**Fig.6a,b** bottom row). Small correlations of beliefs within cliques can still be observed, but they are to be expected even if inference is close to exact because of the predominance of short range connections.

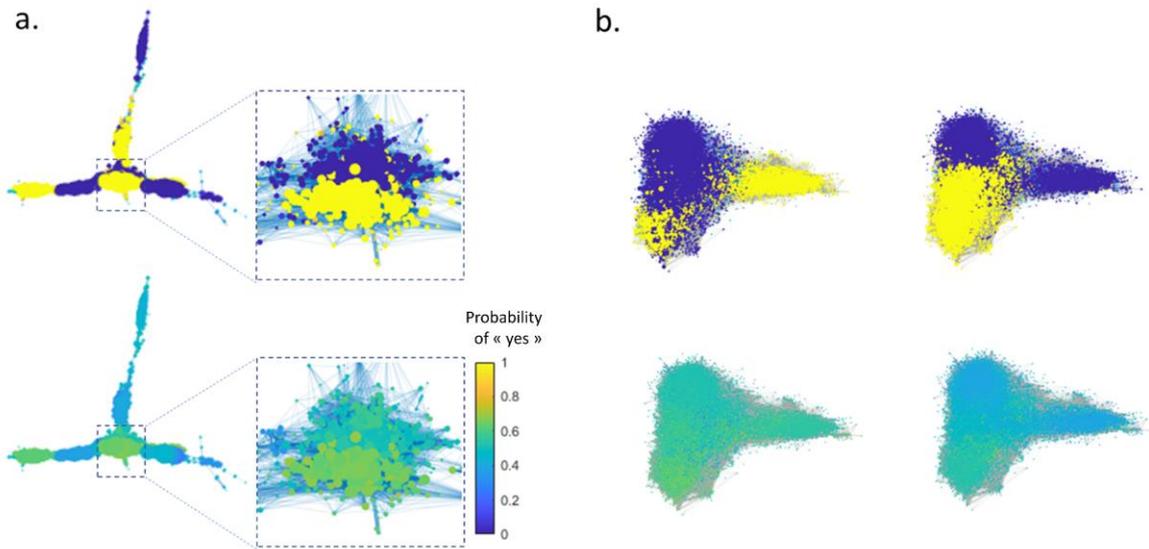

**Figure 6. Social network responses to uniformative external messages**. **(a)** Response of the Facebook© subnetwork with 3959 nodes and 84243 connections, for one example trial, with enlargements of the central part. **(b)** Twitter© subnetwork with 81306 nodes and 1.34 million connections, tested on two different trials (columns). The **top row** represents beliefs computed with BP, and the **bottom row** with CBP. Same legend as in **Fig.3b**.

The more complex structures of these networks made it possible to investigate in more detail the relationships between local graph structures, control parameters (for CBP), and beliefs (**Fig.7**). In the case of BP, overconfidence is directly proportional to the degree of the node being considered (**Fig.7a,b**, *black dots*). Thus, the most connected nodes (agents interacting with many people) develop more extreme views. In contrast, CBP results in far more moderate beliefs and globally weakens (but does not completely remove) the relationship between confidence and node degree (**Fig.7a,b**, *red dots*). CBP achieves this control by learning to decrease the gains ($\kappa_i$) and increase the loop corrections ($\alpha_{ij}$) in nodes of larger degree **(Fig.7c,d)**. In other words, CBP needs to exert stronger controls on nodes that are most massively connected to the rest of the network (influencers) and are thus at the largest risk of becoming radicalized.

To investigate the information sharing capabilities of these networks, we tested them with informative messages provided to small subsets of the nodes, as previously done. In these larger and more modular networks (mean path length 5.5 for Facebook©, 4.9 for Twitter©), both BP and CBP unsurprisingly perform worse than a universal observer (**Fig.8**, left panels). However, CBP strongly outperforms BP. As before, BP exhibits extreme overconfidence, regardless of whether the nodes are congruent (B>0) or incongruent (B<0) in their choices (**Fig.8**, right panels). While the distribution of BP-generated beliefs appears unimodal, it is in fact a consequence of the naturally wide distribution of node degrees in social graphs. If only nodes of similar degrees (e.g., between 20 and 50) are combined, the distribution of belief once again becomes bimodal (**Fig.8c**, left panel), and the separation between the two modes increases with the degree exactly as in **Fig.4b** (for example, imagine measuring the distribution of black dots in a vertical slice in **Fig.7a**). In contrast, the beliefs generated by CBP remain unimodal at all degrees and moderate but with a marked shift and extension toward larger confidence levels as more external information is provided (**Fig.8**, middle panels). In other words, congruent nodes become more confident, while incongruent nodes become less so. Finally, for both BP and CBP, nodes are more likely to be congruent and highly confident if their degrees are larger, that is, if they directly collect messages from a larger portion of the network. This is why the CBP belief distribution not only shifts but also extends to the right as evidence increases.

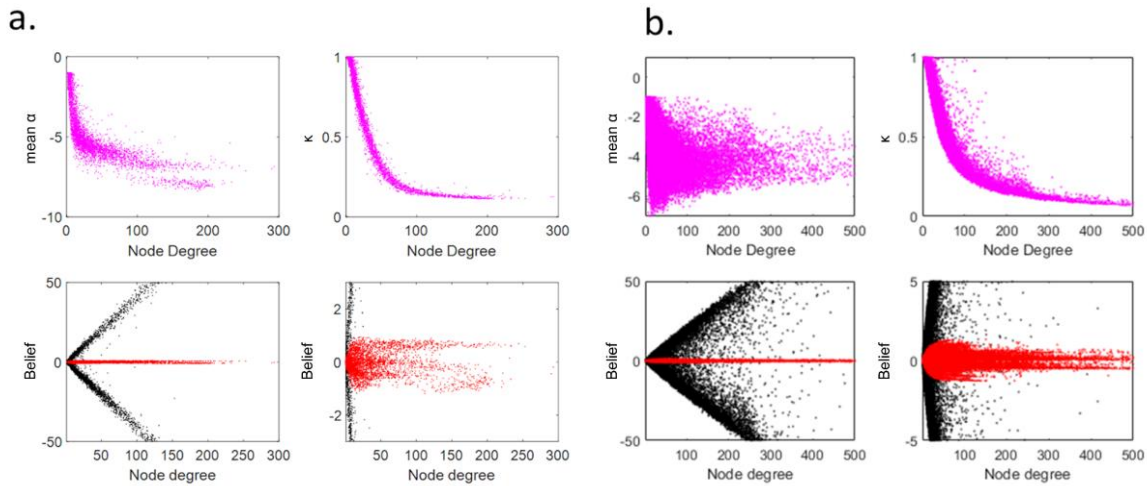

**Figure 7. Social network response to uninformative external messages and learned control parameters as a function of node degree**. **(a)** Facebook© network, **(b)** Twitter© network. Top row: Learned control parameters $\kappa_i$, $\alpha_{ij}$ as a function of node degree. Each dot represents the control parameters for a single node, and the loop correction term is averaged over all incoming connections. Bottom row: Beliefs from BP (black dots) and CBP (red dots) at two different scales of the y axis. The response to a representative trial is shown, with each dot corresponding to one node.

We can predict from these results what would be the consequences of willfully spreading fake news on people's choices and confidence. Both BP and CBP are relatively resilient when it comes to choices: they integrate all the external messages. Fake news would have to overwhelm "real news" to cause a global shift in people's choices. However, the most detrimental effect by far is the potential creation of a small number of extremely polarized nodes, with contrafactual but unshakable beliefs (under BP). This does not take place when reverberation in echo chambers is controlled (CBP). In this case, fake news decreases the mean confidence level but without causing the emergence of extremism (see **Fig.8c**, right vs middle).

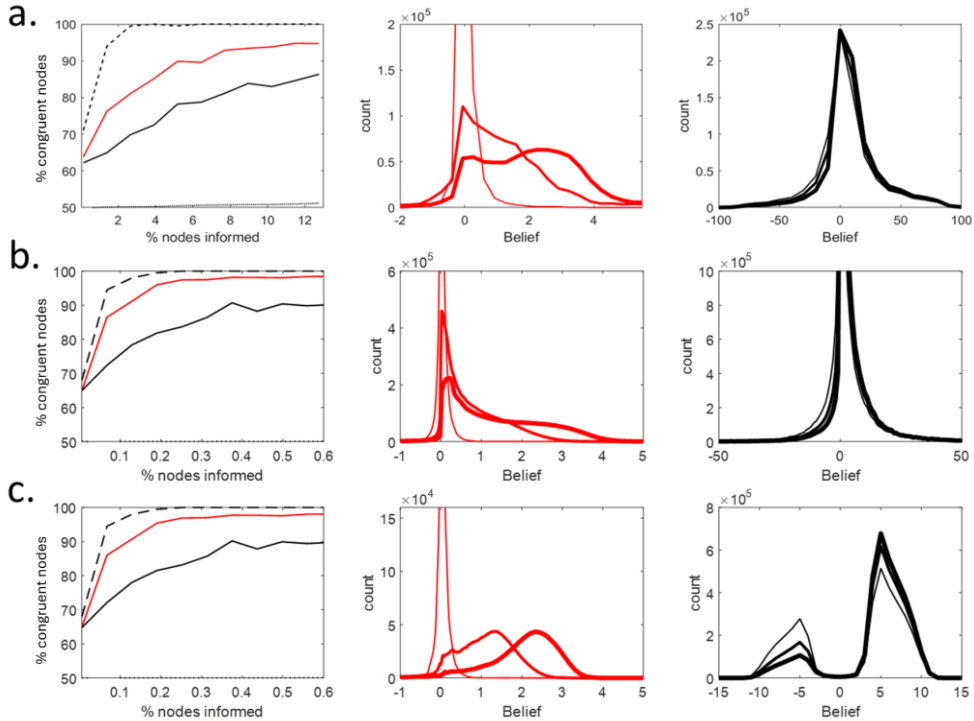

**Figure 8. Response of social networks to informative external messages** (same legends as in **Fig.5**). **(a)** Facebook©. **(b)** Twitter©. **(c)** Twitter© network, but only for nodes with degrees between 20 and 50. The % of nodes informed is unchanged, but the % of congruent nodes is now computed only for this subpopulation. Left panels: Choice performance of the different message passing schemes. Black: BP, red: CBP, dashed: universal observer, dotted: external messages only. Middle panels: Belief distribution over all nodes for increasing amounts of external information as a result of CBP. Right panels: same as middle panel, but for BP. Thin line: 1% of nodes informed, normal line: 5% of nodes informed, thick line: 10% of nodes informed.

## DISCUSSION

Social media networks have always been the theater of repeated questioning and reassessment of ideas that were previously considered unshakable. On the one hand, they have repeatedly demonstrated their invaluable power in bringing together thousands of people to support common important causes of the 21$^{st}$ century. This was notably illustrated by the *MeToo* hashtag, which is famous today for denouncing sexual harassment and abuse, allowing the empowerment of survivors and often forcing societal actions against perpetrators. On the other hand, the way social networks shape public debate has also been exacerbated by populist parties and supporters of conspiracy theories [18]. This last phenomenon appears to directly benefit from real-world uncertainty [15,19] as well as from the viral spreading of information that may reinforce a monolithic (and often extreme) view [20]. Beyond simply modeling echo-chambers in social networks (here, with the Belief Propagation model of communication), algorithms inspired from CBP could provide a solution to moderate overconfidence, going against the effects of echo-chambers and current recommendation systems.

Numerous other theoretical models have been proposed to describe opinion formation in social networks [5,6,21–28]. Some, like the famous voter model, are simple enough to allow complete mathematical analysis [25]. Others have used Ising models [24] or investigated the specificity of small-world social graphs [26]. While many previous models have used binary opinions, others represented them on a continuum of belief [27] such as our model. However, all these models fundamentally differ from ours, not only in their mathematical details, but more importantly in their starting point and objective. All previously cited models take *descriptive* approaches: their starting point is the description of how agents locally interact and their goal is to understand the emergence of collective dynamics. On the contrary, our approach is *normative*: its starting point is a functional hypothesis about the purpose of communication - more precisely, that our opinions are formed optimally when considering the whole external information and the levels of trust between individuals (see [29] for another example of a normative approach, based this time on collective evidence accumulation). Finally, the objective of our approach is to find strategies to achieve this function, or come close to it.

We propose that the root of the radicalization problem is not disinformation or cognitive biases per se (although they certainly play a role). Rather, online message reverberation leads to systematic overconfidence, as information is unknowingly amplified in *echo chambers*. By using a normative approach of opinion formation in a social graph and exploring graphs of progressive complexity, we quantified these phenomena and demonstrated its generality. In popular online social networks (Facebook© and Twitter©), the resulting strength of convictions will largely exceed the available evidence and irremediably lead to the emergence of incompatible world views in different communities (subparts of the network). Confidence levels may become so extreme that opinions are virtually unshakable, remaining the same regardless of the amount of contradictory evidence.

A parametric investigation of Watts-Strogatz graph models lead to further predictions on how specific aspects of social network structure may impact radicalization and polarization. Radicalization was found to depend on the node degree, while being otherwise independent of how connections are organized (**Fig.4**). In contrast, polarization occurred only for small β (i.e. for highly clustered connections). In real social networks, the radicalization of a user should thus increase with its number of active connections (see also **Fig.8**). In contrast, polarization should be more severe, e.g. occur more often and for less controversial topics, in graphs with higher global clustering coefficients. This last prediction, to our knowledge, remains to be tested experimentally.

Borrowing from variational methods of approximate inference in graphs, we proposed that the *Circular Belief Propagation* algorithm (CBP) can alleviate these detrimental effects. This algorithm learns to suppress messages according to how predictable (redundant) they truly are. In small graphs, we showed that CBP achieves close to optimal performance: the social network generates confidence levels that go hand-in-hand with the amount of available evidence. In larger graphs and realistic social networks, CBP avoids radicalization and polarization and ensures that beliefs remain rational.

We are social beings who have exchanged information for millions of years. It would be surprising if we did not have inbuilt cognitive and social strategies to deal with echo chambers in local communities. What may have changed recently is rather the scale and speed of social communication compared to what was previously possible [1]. The worsening trend (increase in radicalization and polarization) appears to be domain-general and applies to many different fields. Beyond the political scene that is often taken as an illustration [30], the scientific community is not immune to overconfidence, particularly since scientific debates have spread from polite but limited academic circles to social media. As a recent example, the results of a trial on the clinical and brain effects of psychedelics in depressive disorders [31] were vividly discussed online with unusual levels of passion even for a scientific debate (see, for instance, this blog entry relating the dispute [32]).

Interestingly, problems that are naturally associated with highly interactive social communication may have their counterpart in the maze of our brain cells, another example of large-scale cyclic graphs. Indeed, the CBP algorithm used here was originally proposed in the context of hierarchical brain structures to investigate reverberations in feedforward/feedback circuits, causing so-called *circular inferences* [33]. Controlling for reverberations in the brain could involve ubiquitous neural mechanisms such as enforcing the excitatory-to-inhibitory balance [34] and account for puzzling perceptual phenomena such as bistable perception [35].

Our simplified model of how communication changes people's opinions does not incorporate numerous aspects of social media communication. For instance, messages are not systematically broadcasted and connections are not necessarily symmetrical (for instance, messages propagate more often from influencers to followers than the reverse). Incorporating this new element into the model would limit polarization. Besides, while we considered stable states after unlimited message exchanges, temporal aspects were ignored. In real life, a piece of news is only propagated for a limited amount of time before becoming obsolete, and our beliefs are constantly updated as new information arrives. On the other hand, we also did not incorporate phenomena that could amplify the severity of echo chambers, such as biased information access (e.g., AI-powered chatbots fastening the spread of fake news [36]), past individual history and priors, or recommendation systems based on preferences [6] that might be used by social media to reinforce the weight of past online activities. Additionally, the present model considers fixed and positive connections, while individuals tend to communicate only with people having similar convictions [37], may distrust others [38], and may even actively distrust people with opposite convictions [39] . This last phenomenon would favor polarization. Lastly, the model only tackles communication over one particular topic, although people form opinions on many questions, and discussion about a subject influences our thoughts on related subjects [9]. Despite these theoretical limitations, going towards an experimental validation of the model would be a giant leap forward. Simple online or offline experiments have been proposed, and could potentially be modeled with either BP or CBP.

Future work will have to determine how the change brought by our proposed model (CBP compared to BP) could be implemented or promoted in real life, as this proposed solution remains theoretical for now. One way would be to inform people by displaying a measure of local polarization caused by the structure of their local interaction graph. This could make users integrate information differently, possibly in a CBP manner. Another way would be to act on recommendation systems by designing them to promote open-mindedness, which could help break echo chambers. This could mean reordering posts on social feeds to propose content according to their unpredictability for the user. This reordering could be monitored by users, for instance through a novelty scale.

# **METHODS**

Here we describe how to reproduce the simulation results. For the theoretical foundation of BP and CBP equations, see the ***Supplementary Material - Math Appendix***.

1. Social graph models

Social graphs were formalized as Ising models with coupling strengths $\{J_{ij}\}$ and biases corresponding to the external messages $\{M_{\text{ext}\to i}\}$. Watts-Strogatz small-world graphs were generated as follows. First, a ring network was constructed by connecting each node to its $K$ neighbors on the right and left. Next, with a probability $\beta$, this local connection was transformed into a long range connection between two randomly selected nodes. The structure of the realistic social networks were obtained from open source data, https://snap.stanford.edu/data/ego-Facebook.html for data from Facebook© and https://snap.stanford.edu/data/ego-Twitter.html for data from Twitter©.

We assumed that coupling strengths were positive (since we communicate with others we trust). For each graph, coupling strengths were selected from a uniform distribution between $0$ and $J_{max}$. We chose the following values in order to maintain reasonable levels of confidence despite increasing graph sizes: $J_{max}$ = 0.6 for 10 node graphs, 0.36 for 200 node graphs, and 0.18 for realistic social graphs. None of the results reported here depends on the exact settings for these coupling strengths.

2. Generation of external messages

*Uninformative* messages (used for training the control parameters and for measuring radicalization/polarization in the absence of meaningful evidence) where sampled independently from a Gaussian distribution with mean equal to 0 and standard deviation $\sigma_{\text{ext}} = 1$ (for small graphs with 10 nodes as in **Fig.2**) or $\sigma_{\text{ext}} = 0.1$ (for bigger graphs with 200 nodes and for realistic social graphs).

*Informative* external messages (see **Fig.5** and **Fig.8**) were sampled independently from a Gaussian distribution with mean +/- 0.05 (the sign defines the so-called *congruent* choice) and equal variance $\sigma_{\text{ext}} = 0.05$. These informative external messages were provided sparsely to only a portion of the nodes $m/n$, where $n$ is the number of nodes in the graph and $m$ corresponds to the number of nodes receiving non-zero external messages (the percentage of informed nodes is $100\frac{m}{n}$). For each value of $m$, we generated 200 sets of informative external messages, each sampled independently from the same Gaussian distribution. Each time, these messages were fed to a different random selection of $m$ nodes. After running the BP or CBP algorithm, we measured the final percentage of nodes with $B_i > 0$, which we called "percentage of congruent nodes". This percentage was averaged over the 200 trials. In the case of the 200 node toy models (see **Fig.5**), this was also averaged over 6 different random networks generated with the same structural parameters K and $\beta$.

3. Message passing algorithms

After being initialized at $M_{i \to j} = 0$, messages were propagated according to a damped version of the update equation provided in the Results section (see also ***Math Appendix***):

$$M_{i \to j}^{t+1} = (1-\tau)M_{i \to j}^t + \tau f_{ij}(B_i^t - \alpha_{ij}M_{j \to i}^t)$$

(Equation 4)

All messages were updated simultaneously for a total of 100 iterations, using $\tau = 0.2$ (the volatility, or rate of forgetting the old information).

The coupling function used in CBP is:
$$f_{ij}(x) = \tanh^{-1}\left(W_{ij}\tanh(x)\right)$$

(Equation 5)

where $W_{ij} = \tanh(J_{ij}) \in [0;1]$ since coupling strengths $J_{ij}$ were taken to be positive. $f_{ij}$ is bounded between $-J_{ij}$ and $+J_{ij}$ and has a sigmoidal shape. Note that function f_{ij} closely relates to the one used in other models (which all consider $\alpha_{ij} = 0$) [5,10], namely, the function $x \mapsto W_{ij}\tanh(x)$.

4. Parameter optimization

Control parameters for CBP were adjusted to the specific graph structure and in order to improve inference as compared to BP (see below). We considered two methods, supervised learning or unsupervised learning.

In *supervised learning optimization* (applied in this work exclusively to graphs with 10 nodes), the exact marginals $p_i(x_i)$ were computed using the junction tree algorithm [40]. The control parameters were chosen using supervised learning to minimize the mean squared error between the exact marginals $p_i(x_i)$ and the ones from CBP $b_i(x_i) = e^{(x_i+1)B_i}/(1+e^{2B_i})$ (where $x_i = 1$ for "yes", $x_i = -1$ for "no") over a set of 300 training examples (trials with uninformative messages):

$$[\alpha^*, \kappa^*] = \arg\min_{\alpha,\kappa} \sum_{\text{Trial}} \sum_i \sum_{x_i} (b_i(x_i) - p_i(x_i))^2$$

(Equation 6)

Control parameters are therefore chosen, using this method, according to an objective function to be minimized. As CBP with these chosen control parameters outperform not only its special case BP on the data it was trained on, but also generalizes well to other external evidence (see **Supplementary material**) unseen during training. The found control parameters depend on the graph couplings in a non-linear, implicit way.

To propose *unsupervised learning* rules (applied to graph with 10 nodes or more), we noted that when the BP algorithm runs on a non-cyclic graph (in which case it performs exact probabilistic inference), messages in opposite directions $M_{i \to j}$ and $M_{j \to i}$ come from completely disjoint parts of the graph and are therefore uncorrelated. The same is true for different incoming messages to the same node (such as $M_{i \to j}$ and $M_{k \to j}$). When external messages (inputs to the graph) are uninformative - and thus uncorrelated -, these internal messages also remain uncorrelated. In contrast, in a cyclic graph, BP results in undue correlations of these opposite messages, which is a direct signature of information reverberation and overcounting in the graph [41]. We thus used, based on these observations, the following explicit (unsupervised) learning rules on control parameters that aim at suppressing these detrimental correlations and ensure that they did not result in spurious belief amplification:

$$\Delta \alpha_{ij} \propto M_{j \to i}(B_i - \alpha_{ij} M_{j \to i})$$ (Equation 7a)

$$\Delta \kappa_i \propto -(B_i^2 - \sum_j M_{j \to i}^2 - M_{\text{ext} \to i}^2)$$ (Equation 7b)

where $\propto$ corresponds to proportionality. We generated 2000 training trials with uniformative external messages, and initialized control parameters as their default BP values $(\alpha, \kappa) = (1, 1)$, before updating them after each trial. Learning rates were chosen separately, to ensure that control parameters properly converged within the training window.

Because coupling weights are positive, the (anti-Hebbian) learning rule for $\alpha$ enforces uncorrelated incoming and outgoing messages $M_{ij}$ and $M_{ji}$. The learning rules for $\kappa$ gain-modulates beliefs according to how strongly incoming and external messages are correlated with each-other, and therefore fights against spurious belief amplifications. Importantly, we checked that these learning rules applied to an acyclic graph converge to $(\alpha, \kappa) = (1, 1)$ which corresponds to the BP algorithm (which is optimal for exact inference in acyclic graphs). This purely heuristic unsupervised approach results in suboptimal inference (see **Fig.2**) but nevertheless, can suppress polarization while improving the information sharing ability of the model social networks.

For both supervised and unsupervised methods, control parameters highly depend on the strength of the graph couplings, to the density of the connections within the graph. The number of nodes within the graph matters less than the local structure of a node (as the number of neighbors per node, and the proportion of neighbors connected to each other).

5. Measures of radicalization of polarization

In **Fig.4**, radicalization was computed by averaging the mean absolute belief $|B_i|$ over all nodes, test trials and network structures (6 randomly generated networks were tested for each combination of $\kappa$ and $\beta$). Polarization was measured as the standard deviation of the beliefs, computed over nodes within a single trial, and then averaged over trials and network structures.